\documentclass{emulateapj} \usepackage{apjfonts}

\shorttitle{The Formation of Fossil Galaxy Groups in the hierarchical Universe}
\shortauthors{D'Onghia, Sommer-Larsen, Romeo, Burkert, Pedersen, Portinari, Rasmussen}

\begin{document}


\title{The Formation of Fossil Galaxy Groups in the hierarchical Universe}


\author{E. D'Onghia\altaffilmark{1,2}, J. Sommer-Larsen\altaffilmark{3}, A.D.Romeo\altaffilmark{4}, 
 A. Burkert\altaffilmark{2}, K. Pedersen\altaffilmark{3}, L. Portinari\altaffilmark{5}, 
J. Rasmussen\altaffilmark{6}}

\altaffiltext{1}{Max-Planck-Institut f\"ur extraterrestrische Physik,  85748
  Garching, Germany; donghia@usm.uni-muenchen.de}
\altaffiltext{2}{Institut f\"ur Astronomie und Astrophysik,
  Scheinerstrasse 1 81679 Munich, Germany}
\altaffiltext{3}{Niels Bohr Institute, Juliane Maries Vej 30, DK-2100 Copenhagen {\O}, Denmark}
\altaffiltext{4}{Osservatorio Astronomico di Capodimonte,
Salita Moiariello 16, 80131 Napoli, Italy}
\altaffiltext{5}{Tuorla Observatory, V\"ais\"al\"antie 20, FIN-21500 Piikki\"o, Finland}
\altaffiltext{6}{School of Physics and Astronomy, University of Birmingham, Edgbaston, B15 2TT Birmingham, UK}


\begin{abstract} 
We use a set of twelve high-resolution N-body/hydrodynamical simulations 
in the $\Lambda$CDM cosmology to investigate the origin and 
formation rate of fossil groups (FGs), which are X-ray bright galaxy groups dominated by a
large elliptical galaxy, with the second brightest galaxy being at least two 
magnitudes fainter. The simulations invoke
star formation, chemical evolution with non-instantaneous recycling, metal
dependent radiative cooling, strong star burst driven galactic
super winds, effects of a meta-galactic UV field and full stellar population
synthesis. 
We find an interesting correlation between  the magnitude gap
between the first and second brightest galaxy and the 
formation time of the group. It is found that FGs have assembled
half of their final dark matter mass already at $z\ga1$, and subsequently
typically grow by minor merging only, wheras non-FGs on average form later.
The early assembly of FGs leaves 
sufficient time for galaxies of $L \sim L_*$ to merge into the central one
by dynamical friction, resulting in the large magnitude gap at $z=0$.
A fraction of 33$\pm$16\% of the groups simulated are found to be fossil,
whereas the observational estimate is $\sim$10-20\%.
The FGs are found to be X-ray over-luminous relative to non-FGs of
the same optical luminosity, in qualitative agreement with observations.
Finally, from a dynamical friction analysis is found that only because infall 
of $L \sim L_*$ galaxies happens along filaments with small impact 
parameters do FGs exist at all.
\end{abstract}




\section{Introduction}
Galaxy groups represent the natural environment where galaxy mergers
are expected to occur most frequently, because of the low velocity 
dispersion of the group galaxies compared to that in galaxy clusters,
as well as the large galaxy number density.
If merging is an efficient process within 
galaxy groups, then there may exist systems in which member galaxies  
merged to form a large elliptical type galaxy, surrounded by considerably
smaller galaxies, for which the dynamical friction infall time exceeds
a Hubble time. In the past a lot of work have been done
to understand the formation and dynamical state and evolution of dense compact groups
(Diaferio et al. 1994; 1995; Hernquist et al. 1995; Governato et al. 1996).
Early numerical simulations of Barnes (1989) showed that compact group members
can merge to form a large elliptical galaxy on timescales considerably less
than the Hubble time.
An observational counterpart to this, called `fossil groups' was discovered 
by the ROSAT X-ray satellite (Ponman et al. 1994). The archetype is 
RXJ1340.6+4018 at $z$=$0.171$ with
a bright isolated elliptical galaxy of magnitude $M_{R}\approx -23.5$ ($h$=0.7)  surrounded by an 
extended halo of X-ray emitting hot gas, and only much fainter satellite 
galaxies.
A fossil group (FG) is defined as a group
where a large elliptical galaxy (BG1) embedded in a X-ray halo dominates 
the bright end of the galaxy luminosity function, 
the second brightest group member (BG2) being at least
2 R-band magnitudes fainter (Jones et al. 2000, hereafter JPF00). 
FGs host nearly all ``field'' E galaxies
brighter than $M_{R}=-22.5$ ($h$=0.5) 
(Vikhlinin et al. 1999, hereafter V99) and have mass-to-light 
ratios comparable to those of galaxy clusters, $M/L_R \sim 300$. 
Their X-ray luminosity is for a given X-ray temperature on average larger 
than for normal groups, such that they appear to fall on the extention
of the cluster $L_X-T_X$ relation to lower temperatures  
(Jones et al. 2003, hereafter J03).
With a number density of $\sim 2$x$10^{-7}$ Mpc$^{-3}$-$4$x$10^{-6}$,
they are not
rare: they constitute $\sim$10-20\%  of all clusters and groups
with an X-ray luminosity greater than $2.5 \cdot 10^{42} h_{50}^{-2}$ 
ergs$^{-1}$ (J03, V99).
The 
luminosity function of FGs shows a lack of L$^*$ galaxies although such groups 
have masses of typically 25\% of the mass of Virgo cluster. Whereas the
Virgo Cluster, e.g., contains six L$^*$ galaxies,
FGs contain typically none, apart from the central, bright E galaxy. 
In terms of the
cumulative substructure function, which is the number of objects 
with velocities greater than a specific fraction of the parent halo's velocity,
FGs like RXJ1340.6+4018 show substructure functions similar to
the Milky Way and M31 rather than, e.g., Virgo (D'Onghia $\&$ Lake 2004).
Hence FGs are excellent probes for testing the present
understanding of the cosmological structure formation, also because 
X-ray observations strongly indicate that they are well relaxed,
virialized systems. 

In this {\it Letter} we present the first attempt, based on self-consistent,
cosmological simulations, to address the physical 
processes that lead to the formation of FGs.

\vspace*{-0.5cm}
\section{THE CODE AND NUMERICAL EXPERIMENTS}

We performed simulations of 12
galaxy group-sized dark matter haloes in the low-density, flat
cold dark matter ($\Lambda$CDM scenario) 
with $\Omega_M$=0.3, $\Omega_{\Lambda}$=0.7, $h=$H$_0/100$
km s$^{-1}$ Mpc$^{-1}$=0.7 and $\sigma_8$=0.9.
The simulations were performed using the TreeSPH code briefly described
in Sommer-Larsen et al. (2005).
The code incorporates 
the ``conservative'' 
entropy equation solving scheme of Springel \& Hernquist 2002; 
chemical evolution  
with non-instantaneous recycling of gas and heavy elements tracing 10 
elements (H, He, C, N, O, Mg, Si, S, Ca and Fe; Lia, Portinari \& Carraro 
2002a,b); atomic radiative cooling depending 
on  gas metal abundance and a redshift dependent, meta-galactic UV field; 
continuing, strong galactic winds driven by 
star-bursts (SNII), 
optionally enhanced to mimic AGN feedback.
A fraction $f_W$ of the energy 
released by SNII explosions goes initially into the ISM as thermal energy,
and gas cooling is locally halted to reproduce the adiabatic super--shell
expansion phase; a fraction of the supplied energy is 
subsequently (by the hydro code) converted into kinetic energy of the 
resulting expanding super-winds and/or shells.

The original dark matter (DM)-only cosmological simulation
was run with the code FLY (Antonuccio et al. 2003), 
for a cosmological box of $150 h^{-1}$Mpc box side length. 
When re-simulating with the hydro-code, baryonic  
particles were ``added" adopting a global baryon fraction of $f_b=0.12$. 
The mass resolution was increased by up to 2048 times, and the force
resolution by up to 13 times.
The initial redshift for the cosmological run was $z_i$=39.  
We randomly selected 12 groups for re-simulation. The only selection
criterion was that the groups should have virial masses close to 
(within 10\%) 1x10$^{14}$ M$_{\odot}$, where the virial mass is the mass at
$z$=0 inside the virial radius, defined as the region for which the average
mass density is 337 times the average of the Universe. The corresponding
virial radius is about 1.2 Mpc, and the virial (X-ray) temperature is
about 1.5 keV. The purpose of this project was to study
a cosmologically representative sample of groups, so no prior information 
about merging histories, was used in the 
selection of the 12 groups.
Particle numbers were about 2.5-3x10$^5$ SPH+DM particles at the beginning
of the simulations increasing to 3-3.5x10$^5$ SPH+DM+star particles at the end.
A novelty was that each star-forming SPH particles of the initial mass is 
gradually turned into a total of 8 star-particles. This considerably
improves the resolution of the stellar component. SPH particles, which
have been formed by recycling of star-particles, will have an eigth
of the original SPH particle mass --- if such SPH particles are formed
into stars, only one star-particle is created. As a result the simulations
at the end contain star-particles of mass $m_*$=3.1x10$^7$  
$h^{-1}$M$_{\odot}$, SPH particles of masses $m_{\rm{gas}}$=3.1x10$^7$ and
2.5x10$^8$ $h^{-1}$M$_{\odot}$, and dark matter particles of 
$m_{\rm{DM}}$=1.8x10$^9$ $h^{-1}$M$_{\odot}$.
Gravitational (spline) softening lengths of 1.2, 1.2, 2.5 and 4.8
$h^{-1}$kpc, respectively, were adopted.

To test for numerical resolution effects one of the 12 groups was in
addition simulated at eight times
(4 for star-particles) higher mass and two times (1.6 for star-particles)
higher force resolution, than the ``standard'' simulations, yielding 
star-particle masses $m_*$=7.8x10$^6$  
$h^{-1}$M$_{\odot}$, SPH particle masses $m_{\rm{gas}}$=7.8x10$^6$ and
3.1x10$^7$ $h^{-1}$M$_{\odot}$, dark matter particle masses
$m_{\rm{DM}}$=2.3x10$^8$ $h^{-1}$M$_{\odot}$, and
gravitational (spline) softening lengths of 0.76, 0.76, 1.2 and 2.4
$h^{-1}$kpc, respectively. For this simulation particle numbers are $\sim$
1.4x10$^6$ SPH+DM particles at 
the beginning of the simulation increasing to 1.6x10$^6$ SPH+DM+star 
particles at the end.
In previous simulations of galaxy clusters (Romeo et al. 2005a,b, 
Sommer-Larsen 
et al. 2005) we have found that in order to get a sufficiently high ICM 
abundance a combination of a large value of $f_W$ and a fairly top-heavy
initial mass function (IMF) has to be employed. We used for the present 
simulations:$f_W$=0.8, an IMF of the Arimoto-Yoshii type. 

\vspace*{-0.5cm}
\section{RESULTS}
For each of the twelve simulated galaxy groups we computed the luminosity
function of the group member galaxies in the B, R, J and K bands 
(Romeo et al. 2005a). 
Our group sample is divided into two classes according to the magnitude of 
the R band gap between the brightest galaxy and the second brightest galaxy: 
FGs ($\Delta m_{12,R}\ge 2$ and  non-FGs ($\Delta m_{12,R}<2$).

A well known problem in the modelling of galaxy groups and clusters,
is the development of late-time cooling flows with bases at the position
of BG1 and associated, central star-formation rates, which are too
large compared to observations. In calculating the optical properties
of the group galaxies we correct in a crude way for this by removing
all stars formed at the base of the cooling flow since redshifts
$z_{\rm{corr}}$ = 2 or 1. Both redshifts correspond to times well
after the bulk of the group stars have formed. The correction amounts
to 20-40\% in terms of numbers of BG1 stars. Using $z_{\rm{corr}}$ =
2 or 1 results in minor differences, so we adopt $z_{\rm{corr}}$=2 in this
paper --- for further discussion of this point see Sommer-Larsen et al. (2005).
In cases where a similar correction of BG2 is appropriate (typically
for non fossil groups, where BG2 enters into the main dark matter halo
fairly late, $z \la$0.5) such a correction is applied to BG2. These
corrections are quite minor, $\la$10\% in terms of numbers of stars.

Figure 1 shows
the composite group luminosity function (LF) for the 4 FGs (triangles) and 
8 non-FGs (pentagons)
of our sample, adopting a bin size of one magnitude.
The large magnitude gap  $\Delta m_{12,R}$  between the brightest galaxy 
and the second brightest galaxy is clear from this figure
for the FGs compared to the absense of a gap for the non-FGs.
The observational LF (shown with stars) is taken from JPF00.
The shape of the LF at the bright end, and the typical number of member
galaxies of our FGs is consistent with the observations to our resolution
limit of $M_R \sim -18.5$ (see below).
Note though, that the LF of RXJ1340.6+4018 (JPF00) is only photometric, so a 
LF with members spectroscopically confirmed 
(Mendez de Oliveira et al. 2005) would enable a more fair 
comparison.
\begin{figure}[t]
\begin{center}
\vspace*{-0.2cm}\resizebox{7.8cm}{!}{\includegraphics{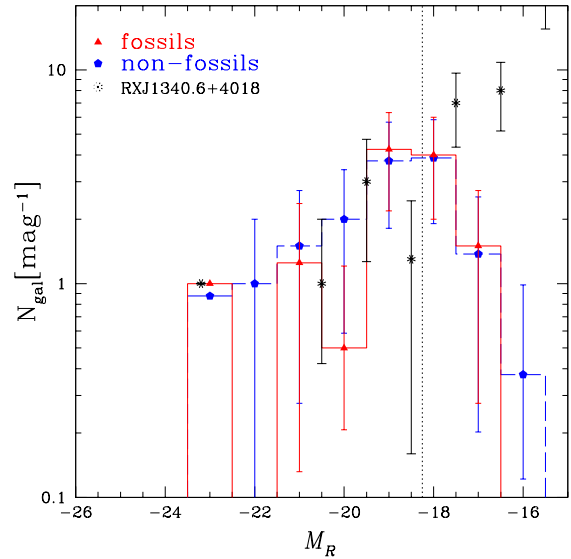}}\
\vspace*{-0.5cm}
\end{center}%
\caption{The composite group LF for the fossil groups (triangles)
 and non fossil groups (pentagons) of our sample. 
The observational LF
for RXJ1340.6+4018 (stars) is shown for comaprison (JPF00).} 

\end{figure} 
In order to test how resolution affects our results we carried out one 
comparison 
simulation at higher resolution (Sec.2). The agreement between the 
luminosity functions of the ``standard'' and higher resolution run is 
excellent above the resolution limit of the normal resolution run, 
$M_R \sim -18.5$, as will be discussed in D'Onghia et al. (2005, paper II - see
also Romeo et al. 2005a for details). 

Figure 2 displays an interesting correlation between formation time of the
group, defined as the epoch at which 50\% of the system's final virial
mass is assembled, and the magnitude gap $\Delta m_{12,R}$ .
This correlation indicates that FGs are  
systems formed at z$\ga$1, and evolved relatively quietly until the present epoch, whereas
non-fossil groups are generally late forming.
The earlier a 
galaxy group
 is assembled, the larger is the magnitude gap in the R band, 
$\Delta m_{12,R}$, at $z$=0. 
\begin{figure}[t]
\begin{center}
\resizebox{7.8cm}{!}{\includegraphics{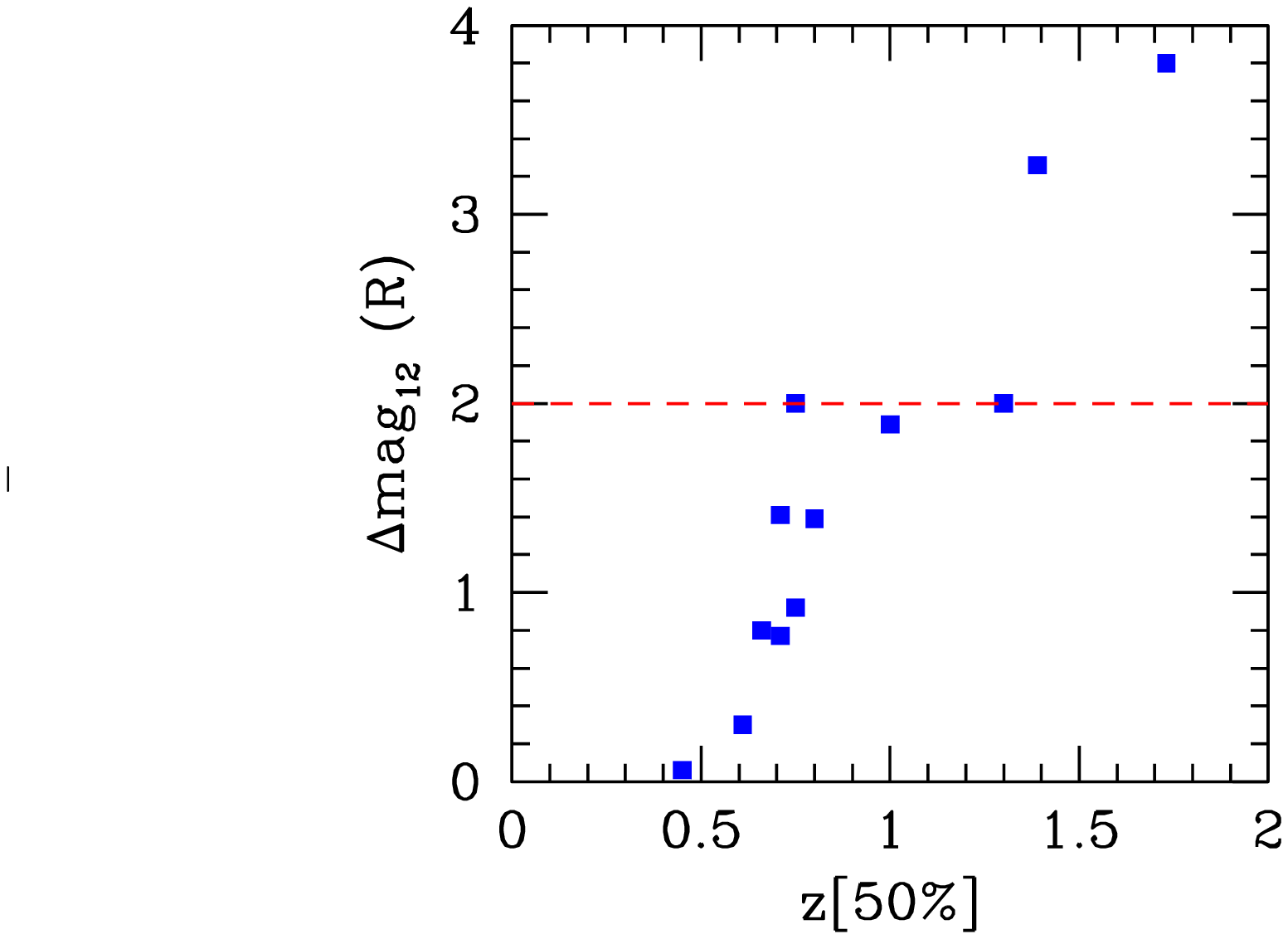}}\
\vspace*{-1.0cm}
\end{center}%
\caption{$\Delta m_{12,R}$ of each simulated group as a 
function of formation time
as defined as the epoch in which the group assembled 50\% of the system's final
mass}
\end{figure}

A characteristic feature of observed FGs is an excess of X-ray luminosity
with respect to ``normal'' galaxy groups of the same optical luminosity.
This feature is reproduced in our simulations:
the FGs with mean optical luminosity L$_R=(1.28\pm 0.06)$x$10^{11}$ L$_{\odot}$
have  L$_{X}=(6.3 \pm 1.1)$x$10^{43}$ erg$\cdot s^{-1}$ 
whereas the non-FGs with L$_R=(1.26\pm 0.04)$x$10^{11}$ L$_{\odot}$
have median X-ray luminosity  
L$_{X}=(1.7 \pm 0.6)$x$10^{43}$ erg$\cdot s^{-1}$ 
 showing a clear offset
between the typical X-ray
luminosity of FGs compared to non-FGs, in qualitative agreement with
observations. 
The reason for this is likely that as the simulated FGs are systems
which form early, gas in the inner regions has more time to cool,
relative to non-FGs, resulting in stronger central cooling flows, and
hence
higher X-ray luminosities. Note, however, that stronger than observed
cooling flows at late times occur in general at the centers of the
simulated groups.
This results in low central gas
entropies and X-ray luminosities about a factor of 3 larger than
observed. 
Similar offsets are found for the L$_{X}$-T$_{X}$ relation of simulated 
galaxy clusters,
when an IMF of the Arimoto-Yoshii type is adopted  (Romeo et al. 2005b).

\section{DISCUSSION}
We have found a correlation between the magnitude gap between the BG1 and BG2 
and the formation time of the dark halo of the groups considered. 
What is the origin of the 
large magnitude gap at the bright end of the LF of FGs?
In Figure 3 we show the evolution with redshift of the stellar mass of  
BG1 and the 
second brightest galaxy at any given time, BG2, for a FG and a 
non-FG from our sample.
At late times ($z\la$1) the ``magnitude'' gap in the FG increases due
to minor merging/stripping.
We interpret this result as follows: 
since the dark halo 
is assembled at $z \ga 1$, there is plenty of time for the 
BG2s to first be tidally stripped during close
passages to the BG1 and finally merge into the BG1, by dynamical
friction.
Note that in Figure 3 for the FG the BG2 is the second brightest
object at each output time. At z=2.2 the BG2  at that time
 merges into the BG1. The last major merger of the BG2 into BG1
occurs at z=1.4 (see arrows in Figure 3). 
\begin{figure}[t]
\begin{center}
\resizebox{8.0cm}{!}{\includegraphics{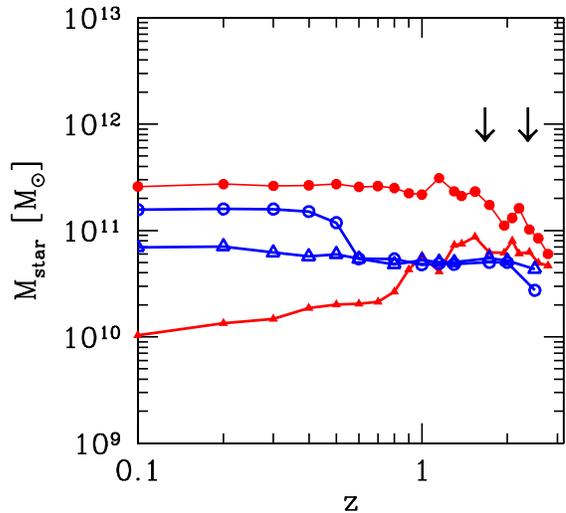}}\
\vspace*{-0.5cm}
\end{center}%
\caption{The evolution of the stellar mass of BG1 (filled circles) and BG2
        at any given time (filled triangles) for a FG. Note that for
        the FG the BG2 is the second brightest object at each output
        time. The BG2 merges into the BG1 at z=2.2 and the last
        major merger of a BG2 into BG1 occurs at z=1.4 (indicated by arrows).
        For a non-FG is
        shown $M_*(z)$ of the galaxy which becomes the final BG1
        (BG2) by open circles (triangles).}
\end{figure}
For the non-FG, the stellar mass of both BG1 (empty circles in Figure 3) and 
BG2 (empty triangles) mostly increases with time till $z$=0. In this case
the final halo assembles its mass fairly recently,
and the two largest, more similar sized galaxies have not had time to
merge.

We now analyze whether the fossil group phenomenon can be explained
by dynamical friction effects. First, we compute the time
$t_{inf}$ it takes for a satellite galaxy moving in a main
dark matter halo to be accreted into the central region of the halo:
we assume singular isothermal sphere models for both the 
main halo and the satellite, with characteristic circular velocities
of V$_{H}$ and  V$_{S}$. If we assume that the satellite initially
moves in a circular orbit (one can show that this gives the minimum infall
timescale for a given initial impact parameter)
at a speed $V=V_{H}$ and initial
radius $r_0$, and that $t_{inf}$ is considerably larger than the dynamical
timescale, the satellite will feel a small torque which slowly reduces its
orbital angular momentum, keeping the orbit almost circular while it 
spirals towards the centre of the halo:
\begin{equation}
\frac{dl}{dt}=V_H \frac{dr}{dt} = a_{\rm{dyn}} r
\end{equation}
denoting $l$ the specific angular momentum of the satellite, $r$ its
orbital
radius and $a_{dyn}$ the drag acceleration opposed to the direction of
motion. The latter can be expressed by (e.g., Binney \& Tremaine 1987):
\begin{equation}
a_{\rm dyn} = \frac{dV}{dt} = - \frac{4\pi \ln\Lambda G^2 m \rho}{V^2} 
\left[\rm{erf}(X) - \frac{2 X}{\sqrt{\pi}} \exp(-X^2)\right] ~~,
\end{equation}  
where $X=V/\sqrt{2}$, $\sigma_H=V/V_H$, $m$ is the mass of the satellite, $\rho$ 
the (mainly dark) matter density, and $\ln\Lambda$ the Coulomb logarithm, 
which from numerical simulations is found to be $\sim1.5$.

The radius of the satellite halo is set by tidal truncation
$r_s=(m/3M)^{1/3}r$, where M is the mass of the main halo
inside of $r$. 
For a isothermal sphere model we have  
$r_s=\frac{V_S}{\sqrt{3}V_H}r$ and 
$m=\frac{V_S^3}{\sqrt{3}V_H G}r$, inserting these equations and eq.(2) in
eq.(1) we obtain:
\begin{equation}
t_{inf}=\frac{r_0}{\frac{dr}{dt}}=12.4 \frac{r_0}{100 \ \rm{kpc}} 
\Big( \frac{V_H}{700 \ \rm{km/s}} \Big)^2  
\Big( \frac{V_S}{250 \ \rm{km/s}} \Big)^{-3} ~\rm{Gyr}~, 
\end{equation}
for $\ln\Lambda$=1.5. Hence, an L$_*$ satellite ($V_S \sim 250$km/s) being
deposited  as close as 100 kpc from the BG1 (given that the group virial
radius at $z$=0 is 1.2 Mpc) within a group of $V_H \sim 700$ km/s 
(corresponding to a virial mass of $10^{14}$ M$_{\odot}$) on a low-eccentricity
orbit has an infall time comparable to the Hubble time. It then at first
seems puzzling that fossil groups exist at all.
However, the infall time is proportional to $r_0$, and as will be discussed
in paper II in the simulations infall happens along filaments with initial
impact parameters as small as 5-10 kpc --- this makes infall times of L$_*$
galaxies less than the Hubble time, and is the main reason why fossil
groups exist. If the early merging history is responsable of the formation
of fossil groups, why are there no clusters with a similar lack of 
$L \sim L_*$ galaxies as a result of cosmic variance? 
As characteristic initial impact parameters scale
linearly with V$_H$, fixing V$_S$ we obtain $t_{inf} \propto
V_H^3$. For a typical rich cluster V$_H \sim 1400-2100$ km/s,
so it follows that the typical infall time for a L$_*$ galaxy
will be 10-30 larger than in groups. Hence massive galaxies have had no time
to merge with the central object yet. 

Given that merging is such an important ingredient in FG formation, we 
check that merger rates are correctly estimated.
When calculating the optical luminosities
of the BG1s (and in some cases also BG2s) we correct for the effect of
late time cooling flows by not including stars formed later than 
$z_{\rm{corr}}$=2  at the base of the cooling flow. These stars will,
however, still contribute to the gravitational field of BG1, which
might lead to an increase of the merger rate. 
To test for this we ran
one simulation, where star particles formed at the base of the cooling
flow (at BG1) were removed entirely from the simulation. 
This did not lead to any significant change of the merger rates.
Moreover, as discussed above, analysis of our high-resolution run indicates
that our simulations are not affected by ``over-merging''.

Our work shows that in the hierarchical structure formation scenario FG halos assemble
earlier then non-FG halos, and they 
preferentially reside fairly isolated in low-density 
environments.  Once the halos assembled 
50\% of their final mass, they evolve 
quiesciently. Consequently, they
experience only little late infall of galaxies
that could stop merging and fill the magnitude gap of the LF again.

We now consider the regions around the two FGs with the  
largest magnitude gaps, $\Delta m_{12,R}>3$, compared to
the regions around the non-FGs: between 2 and 5 Mpc, 
the region around FGs
shows a total matter density in units of the mean of the Universe at z=0,
$\rho_{tot}/\bar{\rho}=2.06 \pm 0.23$, 
whereas the region around non-FG halos has $\rho_{tot}/\bar{\rho}=4.86 \pm 0.99$.
So the region within 5 Mpc around FGs is found to be of lower density (at 
about the 3-$\sigma$ level), 
compared to non-FGs. 
For the region $2 \la R \la 7.5$ Mpc, the corresponding
numbers are $1.88\pm0.39$ and $3.11\pm0.45$, so the region
around the FGs is again underdense with respect to that
of the non-FGs (at about the 2-$\sigma$ level).
From R$\sim$10 Mpc and outwards there is no statistically significant difference between
the environments and the two types of groups. 
This prediction can be observationally tested. If FGs are X-ray emitting systems 
in place  already at z=1, they
might be observed with similar techniques used to discover high-redshift
galaxy-clusters (Mullis et al. 2005). Once the extended X-ray 
source is identified in the XMM-Newton archive, subequent VLT multi-object 
spectroscopy combined with optical/NIR imaging could reveal the possible presence 
of FGs at high redshift.
Furthermore, the correlation function 
of X-ray emitting galaxy groups cross-correlated with galaxies in the nearby 
universe, e.g. from SDSS
data, could be used to test whether FGs preferentially populate low-density 
regions of the Universe. 
 
Of our 12 simulated groups, 4 become FGs corresponding to a fraction of
33$\pm$16\%, larger but still comparable to the observational estimate of 10-20\%, 
although the sample of simulations and observational data is statistically limited.

\vspace*{0.5cm}
We are grateful to the referee for valuable suggestions and to
F. van den Bosch, G. Lake, D. Pierini and T.Ponman for discussions.

\clearpage


\clearpage

\end{document}